\documentclass[sn-mathphys]{sn-jnl}

\usepackage[separate-uncertainty=true,multi-part-units=single]{siunitx}
\usepackage{graphicx}

\jyear{2023}%

\begin{document}

\title[Imaging low-energy positron beams in real-time with unprecedented resolution]{Imaging low-energy positron beams in real-time with unprecedented resolution}

\author[1]{\fnm{Michael} \sur{Berghold}}\email{michael.berghold@frm2.tum.de}
\equalcont{These authors contributed equally to this work.}

\author[1]{\fnm{Vassily} \sur{Vadimovitch Burwitz}}\email{vassily.burwitz@frm2.tum.de}

\author[1]{\fnm{Lucian} \sur{Mathes}}\email{lucian.mathes@frm2.tum.de}

\author[1]{\fnm{Christoph} \sur{Hugenschmidt}}\email{christoph.hugenschmidt@frm2.tum.de}

\author*[1]{\fnm{Francesco} \sur{Guatieri}}\email{francesco.guatieri@frm2.tum.de}
\equalcont{These authors contributed equally to this work.}

\affil*[1]{\orgdiv{Research Neutron Source Heinz Maier-Leibnitz (FRM II)}, \orgname{Technical University of Munich }, \orgaddress{\street{Lichtenbergstr. 1}, \city{Garching bei M\"unchen}, \postcode{85748}, \state{Bayern}, \country{Germany}}}

\abstract{
Particle beams focused to micrometer-sized spots play a crucial role in forefront research using low-energy positrons. Their expedient and wide application, however, requires highly-resolved, fast beam diagnostics.
We have developed two different methods to modify a commercial imaging sensor to make it sensitive to low-energy positrons. The first method consists in removing the micro-lens array and Bayer filter from the sensor surface and depositing a phosphor layer in their place.
This procedure results in a detector capable of imaging positron beams with energies down to a few tens of eV, or an intensity as low as $\SI{35}{e^+ \per s \per mm^2}$ when the beam energy exceeds \SI{10}{\kilo eV}.
The second approach omits the phosphor deposition; with the resulting device we succeeded in detecting single positrons with energies upwards of $\SI{6}{\kilo eV}$ and efficiency up to 93\%. The achieved spatial resolution of \SI{0.97}{\micro m} is unprecedented for real-time positron detectors.
}

\keywords{Positrons, Imaging, Detectors, Semiconductors}

\maketitle

\section{Introduction}\label{introduction}

Low-energy positron beams are an essential tool of investigation in modern material science ~\cite{Mills_PositronSurfaceSciences, Schultz_Surfaces}, enabling a variety of spectroscopic techniques which can be used to study material defects~\cite{Dupasquier_PALSFoundational, Siegel_PALSFoundational}, elemental compositions~\cite{KumarLynn_ElementslDBS}, crystalline surface structures~\cite{Fukaya_THREPD} and Fermi surfaces~\cite{West_ACARFoundation}. The current frontier in the development of positron annihilation spectroscopy technologies is the production and the expedient deployment of micrometer-sized beams, which would allow probing and resolving the properties of a sample to micro-metric scale~\cite{Uedono_Microbeam, Mittender_Microbeam, Gigl_Microbeam}. Furthermore, positron microbeams are an enabling technology for fundamental physics experiments, such as gravitational tests with positronium atoms~\cite{Mills_FreeFall} or interferometry with antimatter~\cite{Sala_Interferometry}.

Due to fundamental technological limitations current low-energy positron beams possess very low brightness.
The difficulty of performing on-line monitoring of a low-energy low-brightness particle beam is a hindrance to the widespread practical application of microscale positron beams. The two most prominently employed techniques to determine a positron beam position and size are the knife-edge method (KEM)~\cite{Platinum} and imaging through the use of micro-channel plates (MCPs)~\cite{Wiza_MCP}. The KEM achieves high transversal resolution but requires long measurement times, since a target has to be moved step-wise through the beam cross-section and data acquisitions performed at each step; furthermore, the resulting measurement is fundamentally uni-dimensional. An MCP, in contrast, allows for fast imaging of the impinging positron beam. Each individual micro-channel in an MCP acts as an amplifier: A charged particle that enters it can trigger the production of a cascade of electrons, which are then converted to visible light by luminescence in a phosphor screen placed in the backplane of the MCP. On the downside, the resolution of MCPs is substantially lower than what is achievable by KEM~\cite{Tresmin_MCP}; furthermore, their installation imposes stringent geometric constraints, such as the availability of a view-port for imaging of the phosphor screen via a digital acquisition camera~\cite{Vinelli1}.

A possibility hitherto unexplored is to skip the amplification stage of the MCP and instead directly impinge a continuous positron beam onto a phosphor-coated image sensor. Belief in the potential of this configuration is in part motivated by the experimental observation that a phosphor screen impacted by a positron beam emits significantly more light than when it is hit by an electron beam of similar energy and intensity; effect that is more pronounced at lower energies~\cite{Stenson_Luminescence}. On top of that, this new approach to beam imaging would exploit the huge progress made in the image sensor technology in the last 30 years: spurred by the popularization of digital cameras and smartphones, the industry has achieved orders of magnitudes of improvements in terms of pixel count, resolution, sensitivity, noise and cost~\cite{Boukhayma_CMOS, Gouveia_CMOS}. The sensitivity of modern image sensors might realistically be sufficient to detect positron-induced luminescence without the need for an electron-cascade amplification stage. Moreover, current image sensors have pixel sizes in the order of \SI{1}{\micro\meter} or smaller, a resolution surpassing that of the best MCPs and comparable to what can be achieved with KEM.

Even more ambitiously, one could attempt to directly use an image sensor without phosphor coating to image and impinging positron beam. The idea of using image sensors for beam detection is in itself not completely new, as the phenomenon of electron beam-induced current (EBIC) can be employed for imaging of electron beams~\cite{Hanoka_EBIC, Mecklenburg_EBIC}. The main challenge here is that the intensity of a positron beam is typically between $7$ and $14$ orders of magnitude smaller than that of an electron beam. A priori, we would have expected that necessary requirements to directly image a positron beam with an image sensor would be the use of a custom built sensor, long integration times and a high intensity positron beam, such as that produced at the NEPOMUC facility~\cite{NEPOMUC}. 

In this work we report that a consumer-grade image sensor, based on complementary metal-oxide substrate (CMOS) technology can, instead, resolve individual impinging positrons with almost unit efficiency and micrometric accuracy if the positron kinetic energy exceeds \SI{6}{\kilo\eV}. This result far surpasses our expectations in terms of sensitivity and resolution. Furthermore, we determined that it is possible to image the beam also by coating the image sensor with phosphors; our results indicate that, in this configuration, it is possible to detect a positron beam with a kinetic energy down to a few \SI{}{eV} and with a resolution that is only limited by the thickness and granularity of the phosphor coating.

\section{Results}\label{results}

\subsection{Phosphor-aided detection}\label{phosphor}

\begin{figure}[ht]
\centering
\includegraphics[width=0.8\textwidth]{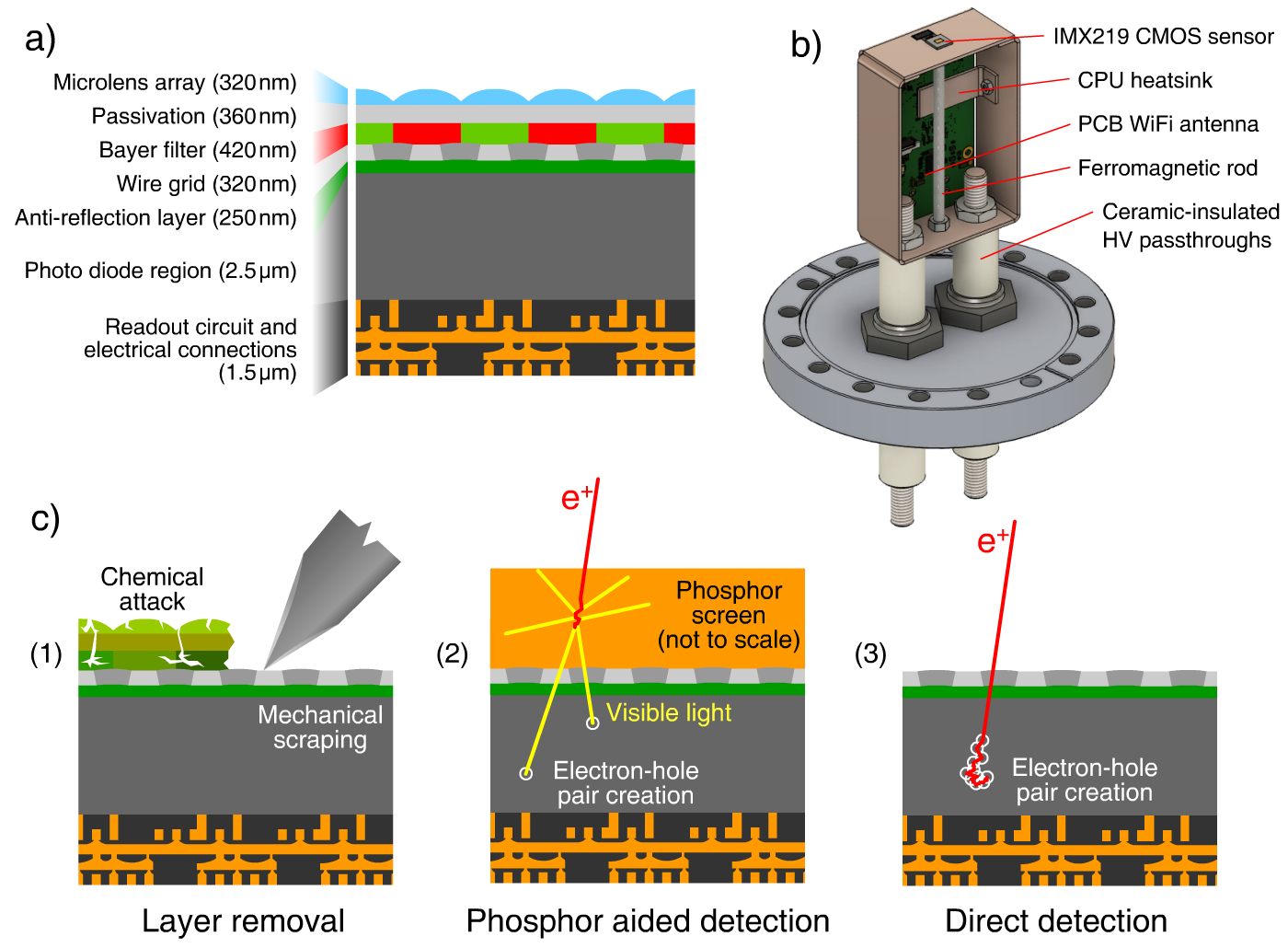}
\caption{a) Section of the microscopic structure of the sensor according to Matthews et al.~\cite{ReverseEngineering} b) Main assembly used to hold and test the sensor; for clarity the front cover of the copper encasing holding the readout electronics is omitted. 
c) Schematic representation of the processes of layer removal (1), phosphor-mediated imaging (2) and direct imaging (3).}
\label{Apparatus}
\end{figure}

We have imaged a low-energy positron beam by means of detectors realized by modifying commercial CMOS image sensors (Sony IMX219). The sensors were modified by first removing the Bayer filter and microlens array from their surface and then depositing a \SI{70}{\micro m} coating of EJ-600 ZnS:Ag phosphor powder (see Fig.~\ref{Apparatus} and section \ref{MethodsSample}).

The beam was produced by moderating the positrons emitted by a radioactive ${}^{22}$Na $\beta^+$-emitter through a \SI{1}{\micro m} thick mono-crystalline (110) tungsten foil. This beam can be focused onto the target in a $2 - \SI{3}{\milli m}$ spot with an energy that can be varied between 2 and \SI{25}{\kilo eV}; positron rate at the target is in the order of $6 \cdot 10^4$ positrons/s  ~\cite{LabBeam}. A variety of coils is used to adiabatically transport the beam, including several pairs of steering coils which allow us to move the spot laterally with respect to the target. The image sensor was installed in the focusing point of the positron beam with its surface orthogonal to the beam direction. Beam focusing is achieved by the combined action of electrostatic lenses and by pinching the magnetic field in the proximity of the target by means of a ferromagnetic rod installed underneath the sensor (see Fig.~\ref{Apparatus}b). 

\begin{figure}[ht]
\centering
\includegraphics[width=\textwidth]{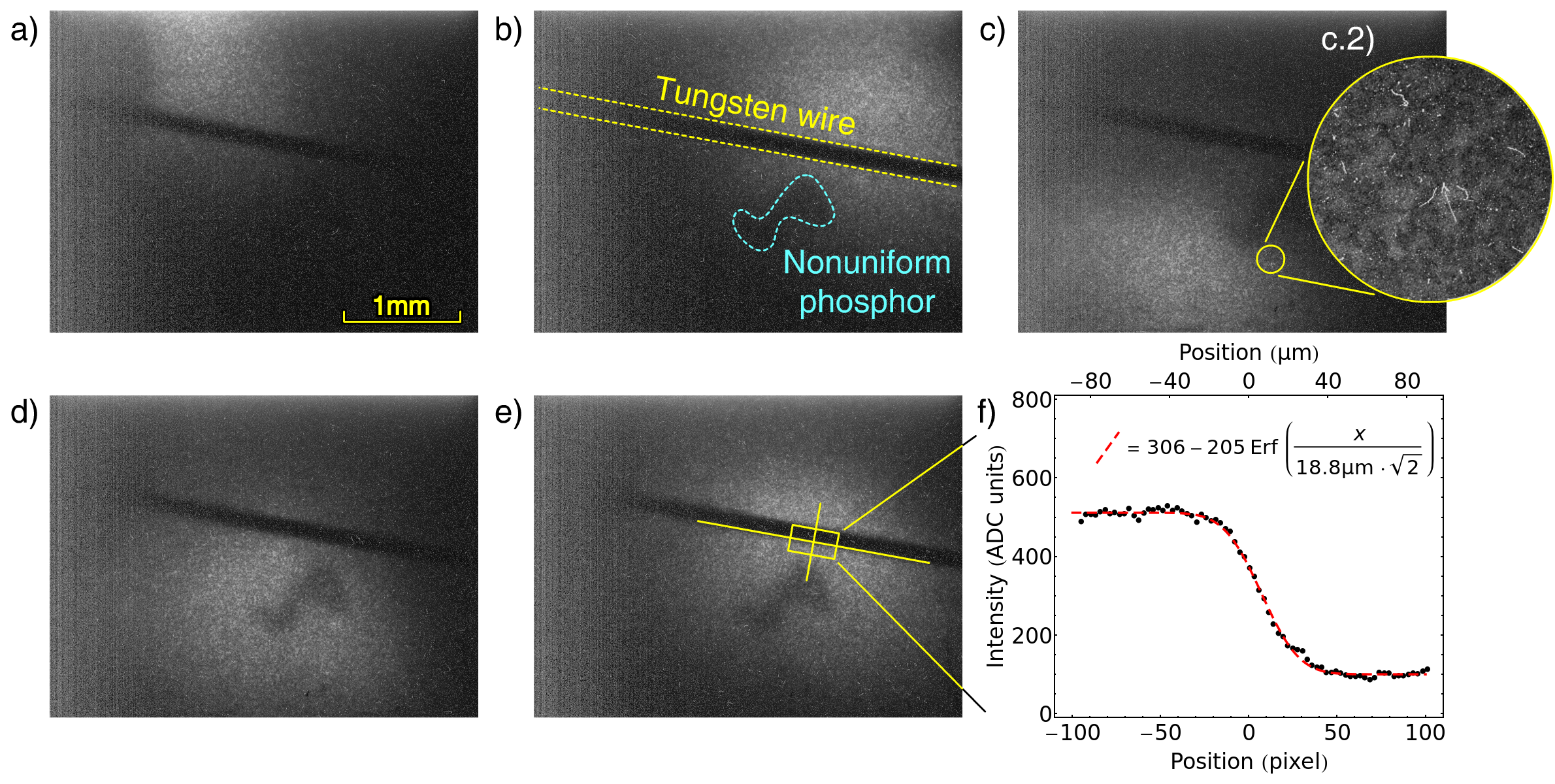}
\caption{a-e) Imaging of the positron beam using the phosphor-coated CMOS sensor with different steering coils settings. The black shadow is produced by a tungsten wire, the dark area below is a region with thicker phosphor coating. c.2) The magnification shows the outline of single phosphor grains and white tracks left by Compton electrons. f) The portion of image e) marked by the yellow rectangle has been projected and fitted with the error function to provide a lower limit to the sensor's OTF; in practice, factors other than the sensor resolution contribute to the blurring, such as beam divergence and scattering over the wire edges.}\label{CoatedImaging}
\end{figure}

The particle beam is recorded by the sensor as a bright spot (see Fig.~\ref{CoatedImaging}). We have verified the nature of the signal to be the positron beam with three independent methods. Firstly, by observing the disappearance of the signal when a shutter located \SI{3}{m} upstream of the sample holder is closed. Secondly, by observing that a tungsten wire with a diameter of \SI{125}{\micro m} installed \SI{1}{\milli m} in front of the sensor surface will cast a shadow onto the sensor and, thirdly, by observing that the spot recorded by the sensor moves in accordance with the variation of the current in the correction coils used to steer the beam (see Fig.~\ref{CoatedImaging}). 
In the image recorded by the sensor it is possible to discern the outline of single phosphor grains, as well as the presence of curly white streaks with lengths ranging from $5$ to \SI{70}{\micro m}. These white streaks are tracks produced by high energy electrons resulting from Compton scattering with high energy $\gamma$ quanta, which themselves are the result of positron annihilation (see Fig.~\ref{CoatedImaging}c.2). The process has a low enough cross section to not interfere significantly with the measurement.
Based on the blurring of the shadow projected by the wire we can determine an upper limit to the standard deviation of the optical transfer function (OTF) of the sensor~\cite{Reichenbach_Resolution} finding it to be less than \SI{19}{\micro m}, which is similar to the resolution given by an MCP and substantially better than what a typical MCP-phosphor-camera system can provide~\cite{Tresmin_MCP}.

\begin{figure}[ht]
\centering
\includegraphics[width=0.7\textwidth]{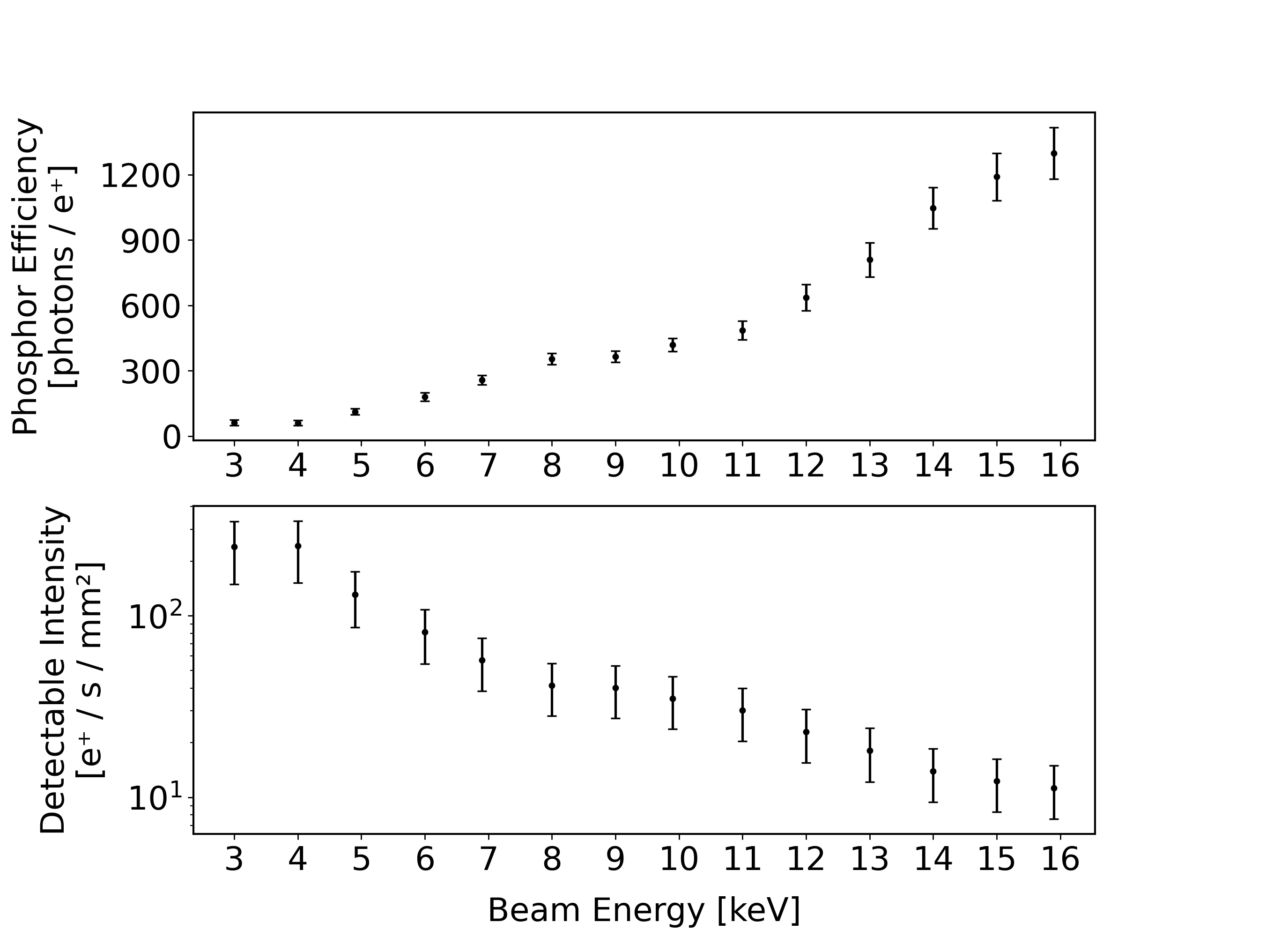}
\caption{Top: the amount of photons emitted by the phosphor coating determined from the sensor as a function of the beam energy  of the signal produced into the image sensor. The sensor was operated with variable gain and exposure time in order to keep the signal within the sensor's dynamic range. Bottom: the minimum beam intensity necessary to produce a detectable signal with a S/N ratio larger than $1$ using an exposure time of \SI{167}{s} and employing $16\!\times\!16$ pixel sub-sampling.}\label{CoatedSignal}
\end{figure}

The beam intensity required to produce an image depends on the exposure time and on the beam energy, as expected from Stenson et al.\cite{Stenson_Luminescence}. We have computed the ratio between the signal recorded by the sensor's analog to digital converter (ADC) and the positron flux in two steps. First we measured the total beam intensity with two high purity Germanium (HPGe) detectors in anti-collinear geometry. We then determined the positron flux impinging on different regions of the sensor by performing a Gaussian fit of the spot (see section \ref{MethodsFit}). Fitting the beam is necessary to prevent the clipping due to the limited sensor area from biasing the calibration of the beam intensity. The relationship between positron flux and recorded signal intensity is shown in Fig.~\ref{CoatedSignal}. When operating the beam with a kinetic energy of \SI{10}{\kilo eV} a positron flux greater than ${35\pm11\,\SI{}{particles/s/mm^2}}$ is detectable with a signal to noise (S/N) ratio greater than 1; to achieve this S/N ratio an exposure time of \SI{167}{s} and $16 \!\times\! 16$ pixel sub-sampling are required. Alternatively, a \SI{19}{s} exposure time combined with $4\! \times\! 4$ pixel sub-sampling will still be able to detect a flux as low as $730\pm360\ \SI{}{particles/s/mm^2}$ allowing for a faster and spatially more resolved measurement. In general, the operational parameters of the sensor should be adjusted depending on the flux to be detected to achieve optimal performance or a specific S/N ratio.
With the beam intensity available to us we are still able to detect the beam when a \SI{2}{\kilo eV} implantation energy is employed. According to literature~\cite{Stenson_Luminescence}, this positron flux should be visible down to an implantation energy of a few \SI{}{eV}.

\subsection{Direct positron detection}\label{direct}

\begin{figure}[ht]
\centering
\includegraphics[width=0.9\textwidth]{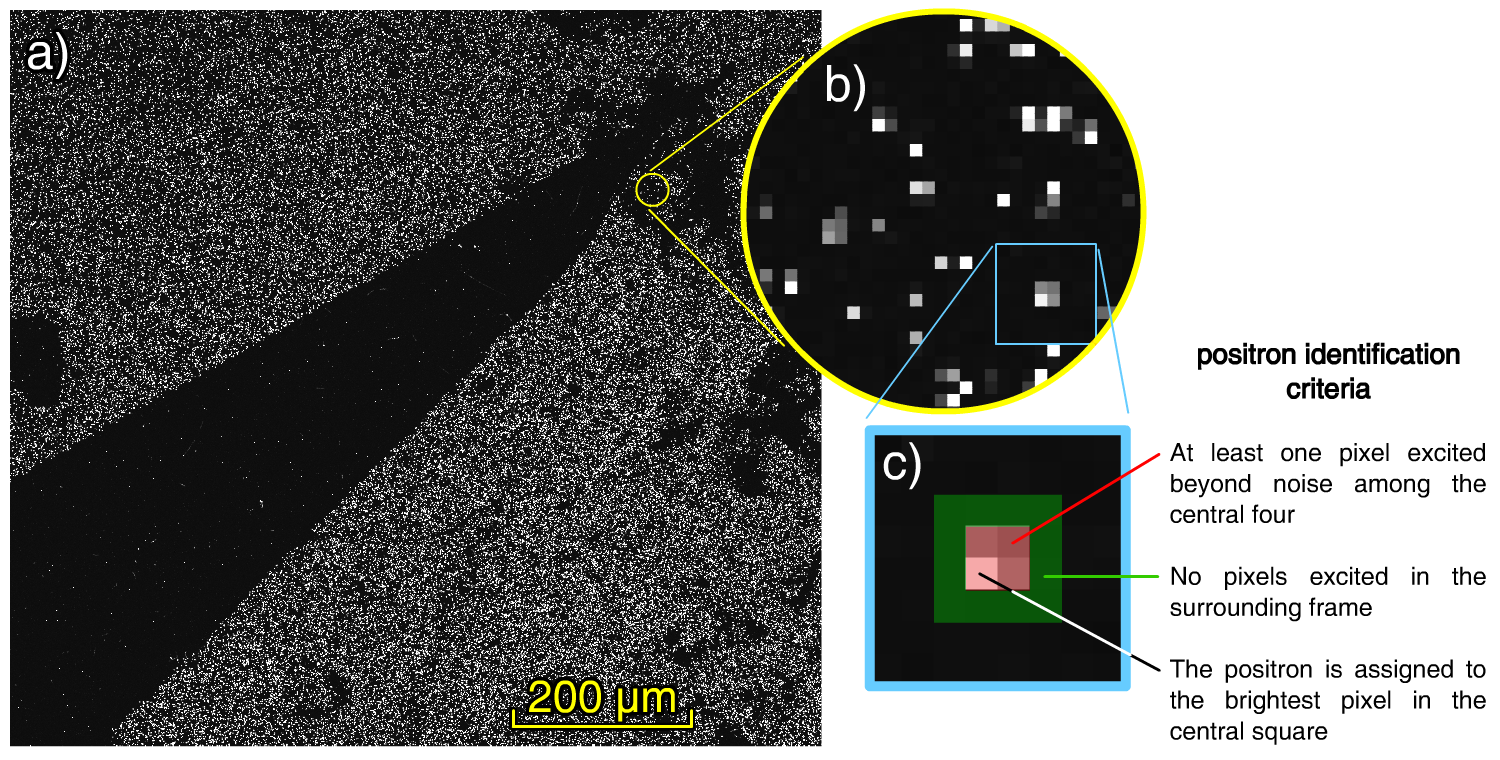}
\caption{a) High resolution imaging of a wedge of aluminum foil obtained with \SI{10}{\kilo eV} positrons impinging on an uncoated CMOS sensor. b) A detail showing instances of the signal produced by single positrons impinging on the surface of the sensor, which results in the activation of one to four pixels. The dark shadows near the tip of the aluminum foil are caused by impurities that were present in the drop of isopropyl alcohol used to deposit the aluminum foil onto the sensor. c) Criteria used to identify positrons in the images.}\label{Cluster}
\end{figure}

We tested whether the CMOS sensor itself is sensitive to low-energy positrons. We prepared the sensor by removing the Bayer filter and microlens array from its surface, but applied no coating (see section \ref{MethodsSample}). The sample was installed in the same setup employed in \ref{phosphor} and tested with the same beam. Acquisition was performed with an exposure of \SI{4.9}{s} and a gain value of $5.3$ while the beam energy was varied between \SI{3}{\kilo eV} and \SI{15}{\kilo eV}. We found the sensor to be, in this configuration, capable of detecting single positrons with kinetic energy larger than of \SI{6}{\kilo eV}. Single positrons appear in the image as clusters of one to four pixels whose brightness greatly exceeds the background noise (see Fig.~\ref{Cluster}).

\begin{figure}[ht]
\centering
\includegraphics[width=0.7\textwidth]{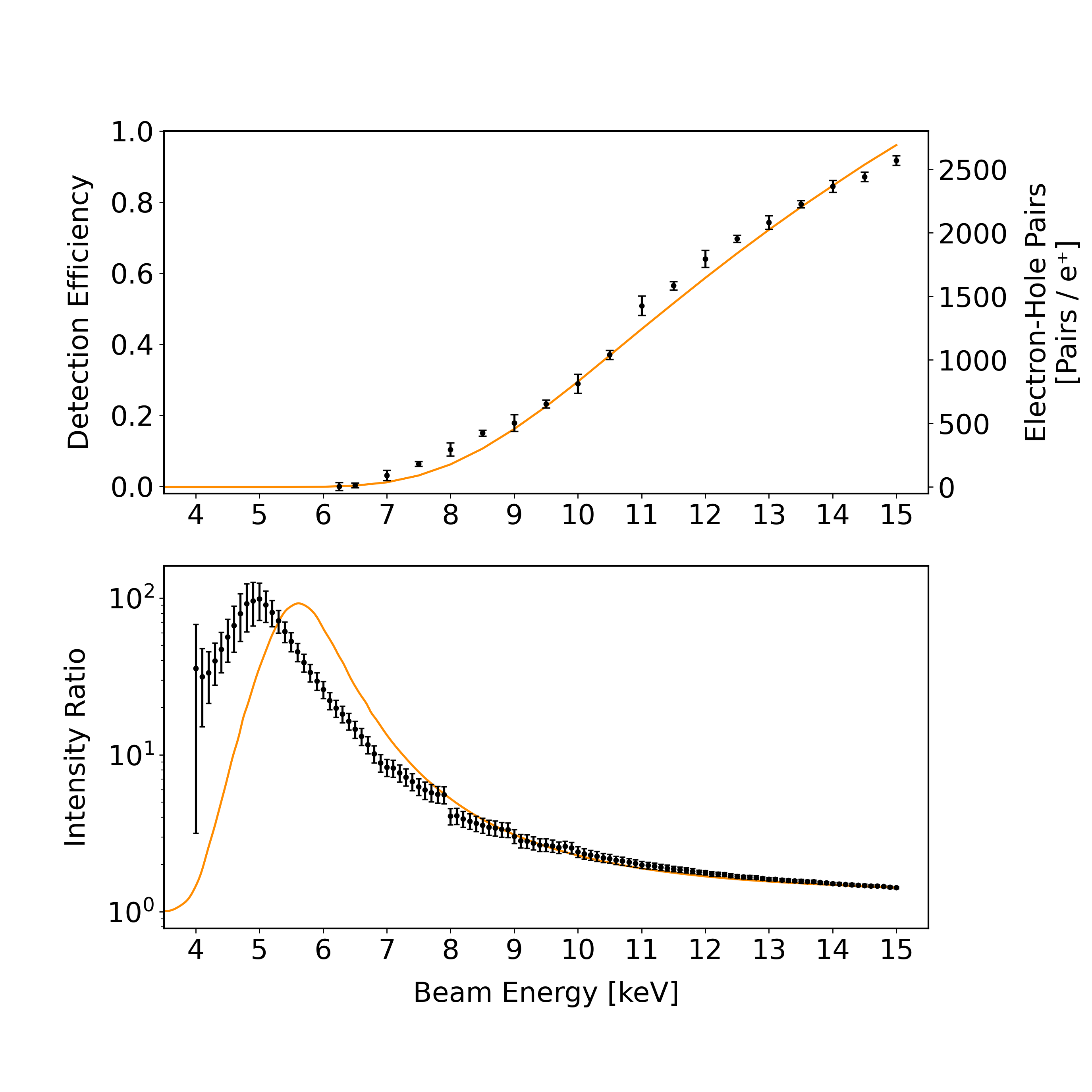}
\caption{Top: Positron detection efficiency of an uncoated CMOS sensor with the Bayer filter and microlens array removed as a function of the impinging positron energy. The detection efficiency is close to unity at \SI{15}{\kilo eV} and drops to zero below \SI{6}{\kilo eV} due to the positron kinetic energy being insufficient for them to reach the sensitive volume of the sensor. The electron-hole pair creation in the sensitive volume of the sensor as predicted by Monte Carlo simulation is shown as orange line. Bottom: the ratio of the signal recorded by a small area of the sensor in which the wire grid was removed and by an adjacent area with the wire grid intact. In solid orange the same ratio as predicted by Monte Carlo simulation.}
\label{UncoatedEfficiency}
\end{figure}

Computing the detection efficiency of the sensor requires knowledge of the positron flux impinging over the sensor surface. This cannot be determined solely from the counts of the HPGe detectors installed around the sample chamber since, as can be seen in Fig.~\ref{CoatedImaging}, the beam spot is not completely contained in the sensor area. We have determined the local positron flux over the surface of the sensor by fitting the beam shape to determine the portion of beam hitting the sensor surface.
To produce a suitable image to perform the beam fit an exposure time of \SI{4.9}{s} and gain of 5.3 were employed, as these parameters produce negligible pile-up and keep the signal within the dynamic range of the pixels. We observed the detection efficiency to depend strongly on the beam energy, ranging from being close to zero for energies below \SI{6}{\kilo eV} to approaching unity for energies above \SI{15}{\kilo eV} (see Fig.~\ref{UncoatedEfficiency}, top). We interpret this result as an indication that the photodiodes built into the CMOS sensor are indeed extremely sensitive to positrons, but that a kinetic energy of \SI{6}{\kilo eV} is required for the positron to penetrate deep enough to reach them. This interpretation is coherent with the available information about the geometric structure of the sensor~\cite{ReverseEngineering}, which indicates the presence of additional layers on top of the sensitive volume, specifically an anti-reflection coating, a wire grid layer and possibly a passivation layer (made of silicon oxide, silicon nitride or both~\cite{Gouveia_CMOS}); we present in section~\ref{MonteCarlo} a quantitative discussion of the effects of these layers.

It is reasonable to assume that the removal of additional layers would allow the detection of positrons of even lower kinetic energy. We have attempted to scrape away the wire grid from the surface of the sensor, but found the process to be too unrealiable to be employed in any practical sense as removing large portions of it almost inevitably leads to the destruction of the device. We were nonetheless able to scrape away a small portion of the wire grid without damaging the sensor. As the scraped  area is too small to allow for a beam fit, it is not possible to compute its absolute detection efficiency; instead we computed, as a function of the implantation energy, the ratio between the signal intensity recorded in this area and in an adjacent area still coated by the grid (see Fig.~\ref{UncoatedEfficiency}, bottom). We removed the fixed-pattern noise (see section~\ref{MethodsElectronics}) from the measurements before computing the ratio, but not the constant pedestal that arises naturally when applying the positron beam. This is needed, as its presence keeps the ratio finite at all energies; for consistency the same pedestal was included in our simulations of this metric (see section~\ref{MonteCarlo}). The ratio shows a peak located around \SI{5}{\kilo eV} with a half width of about \SI{1}{\kilo eV}, indicating a decrease of the energy threshold for positron detection by about \SI{2}{\kilo eV} in the exposed areas.

\begin{figure}[ht]
\centering
\includegraphics[width=0.7\textwidth]{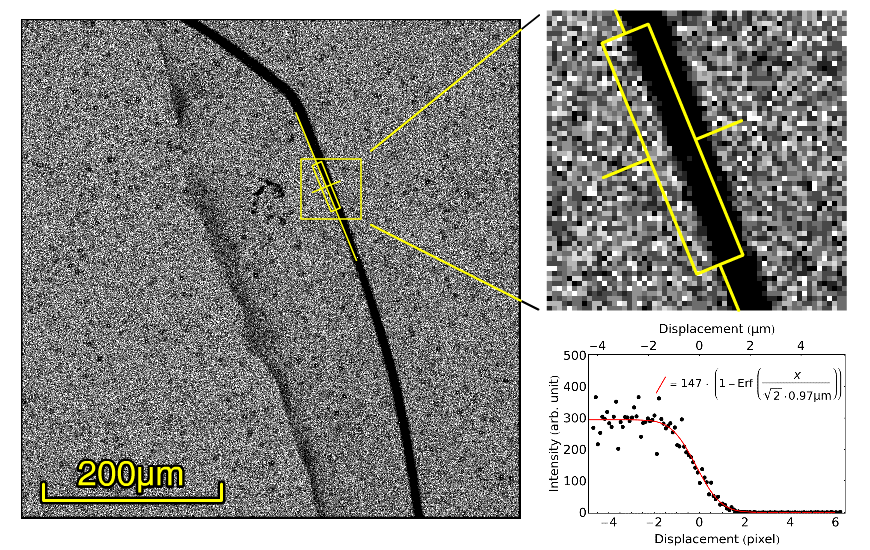}
\caption{Shadow of a \SI{4}{\micro m} tungsten wire laid directly onto the sensor surface. The image was produced by tallying together 7096 images shot with a \SI{4.9}{s} exposure; positrons were identified and their position assigned to one of the image pixels. The gray level in the image indicates the amount of positrons collected within each specific pixel. On the right, fit of the projection of the wire shadow with the error function indicating an OTF with a standard deviation of \SI{0.97}{\micro m}}\label{UncoatedResolution}
\end{figure}

We have determined the resolution of the sensor by imaging the shadow projected by a single strand of tungsten wire, \SI{4}{\micro m} in diameter placed directly onto the sensor surface. The imaging was performed by acquiring 7096 frames with a \SI{4.9}{s} exposure, which is low enough to distinguish individual positrons in the images. Positrons were identified in each of the acquired images as instances of a $2\!\times\!2$ squares of pixels containing at least one whose brightness exceeded the background noise, and surrounded by a 1-pixel thick frame containing no pixels excited beyond the background noise; the positron was then assigned to the brightest pixel among the four (see Fig.~\ref{Cluster}c for a visual representation of the criteria).
We present in Fig.~\ref{UncoatedResolution} the result of the reconstruction in which we determined the OTF width by projecting a portion of the wire shadow and fitting it with an error function~\cite{Reichenbach_Resolution} with standard deviation $\SI{0.97}{\micro m}$.

\subsection{Durability of the sensor}\label{durability}

In our most extensive testing we have implanted a total of $1.2 \cdot 10^{10}$ positrons into one of our detectors; as we did so we observed two ways in which damage to the sensor did manifest. Around $12$ pixels per million behaved as dead pixels and always read out at the upper limit of their dynamic range. An additional $0.28\%$ of the pixels manifest an increase of their dark current; modest in most cases (less than 8\% of the full dynamic range in 75\% of cases) but dramatic in a few instances. By comparing data recorded over the course of 45 hours we observed a slight upwards trend in the dark current of these few pixels, with typical rates in the order of $0.17\%$ of the dynamic range per day. The slow rate at which the dark current increases and the absence of abrupt changes in any pixel's dark current indicates that these pixels were most likely damaged during the sample preparation; furthermore, that the dead pixels were most likely produced in a similar fashion, albeit with a baseline shift severe enough to cause them to saturate at all times. Within the 705 cases of pixels with an anomalous dark current that we have analyzed, less than 10\% displayed a dark current greater than one third of the ADC dynamic range, and as such the anomaly can be in most cases corrected algorithmically; this leaves about 300 pixels per million which are unusable or barely usable. Due to their limited number, we do not expect this kind of damage to be an issue for the vast majority of applications.

\subsection{Monte Carlo simulation}\label{MonteCarlo}

The physical mechanism allowing the direct detection of positrons impinging on a CMOS sensor is the creation of electron-hole pairs inside the depletion zone of the photodiodes built into the device~\cite{Hanoka_EBIC}; this creation occurs through a combination of bremsstrahlung and elastic scattering as the positrons are stopped by the material. We have adapted the code base from a previous work~\cite{NCPImplantation} and employed the available information on the geometric structure of the Sony IMX219 sensor~\cite{ReverseEngineering} to simulate the generation of electron-hole pairs in the sensitive portion of the image sensor. The core of our code simulates the process of positron scattering in silicon as a series of discrete interactions. Silicon is the most abundant element in a CMOS sensor and a suitable stand-in for the other materials that we expect to be present (aluminum, silicon, silicon oxide and silicon nitride) on account of having a similar density. We compute the expected amount of electron-hole pairs created in each discrete interaction as proportional to the energy lost by the positron, weighted depending on the positron energy according to the data presented by Funsten et al.\cite{Funsten1}. The positron detection efficiency is determined as proportional to the expected amount of electron-hole pairs generated in the sensitive region. Placing the sensitive region at the depths presented by Matthews et al.~\cite{ReverseEngineering} ($0.57 - 3\SI{}{\micro m}$) yields a detector efficiency in excellent agreement with the experimental data (see Fig.~\ref{UncoatedEfficiency}).

We used the same simulation process to predict the signal ratio between areas covered by the wire grid and areas not covered by the wire grid, where the sensitive volume lays \SI{250}{\nano m} below the surface. The resulting curve resembles the experimental data, albeit with a shift in energy of about \SI{800}{eV} which cannot be resolved by modifying the depth of the sensitive volume. There are several possible causes for this shift. Firstly, in our simulation we are using silicon as a stand-in for every material present in the sensor: this approximation works best when the sensitive area is deeper and coated by different layers, as the deviations from the silicon model tend to average out. After removing the wire grid only the anti-reflection layer remains, which makes the simulation result more sensitive to its modeling. In addition, it is known that energy deposited in layers adjacent to a p-n junction can result in the creation of electron-hole pairs in the junction itself~\cite{Funsten1}; we can expect this phenomenon to become more prominent in the case of a shallower implantation as it causes the density of deposited energy to increase.

As positrons enter the material and lose energy they also spread transversely and produce a signal that can be picked up by other nearby pixels. We have determined via Monte Carlo sampling that the lateral spread in the active region previously quoted is a distribution with \SI{340}{\nano m} standard deviation. The discretization to pixel coordinates intrinsically introduces an OTF having \SI{323}{\nano m} width. Combining these two yields an OTF with a standard deviation of \SI{470}{\nano m}, which indicates that the resolution found in subsection \ref{direct} is likely overestimating the sensor's actual OTF by about a factor of $2$. On top of that, reconstruction algorithms capable of circumventing the pixel discretization might have the potential of reaching the resolution of the best positron detectors ever built~\cite{Anzi_2020}.

\section{Discussion} \label{discussion}

We have established that modern commercial CMOS sensors achieve, with minimal modifications, exceptional detection performance of low-energy positron beams: they can provide high readout frequency, high resolution and two-dimensional imaging. The imaging can be achieved via conversion of the positrons to visible light in a phosphor coating layer applied onto the sensor; the resulting light is intense enough to be detected without amplification even for weak, radioisotope-based beams. Alternatively, detection can be achieved by implanting positrons with sufficient kinetic energy directly into the sensor.

As an outlook, the use of CMOS technology could usher in the development of a new class of custom detectors that are very precise, sensitive and affordable. Furthermore, they could be designed to integrate the readout electronics within the silicon die, reducing the amount of additional vacuum-installed components and thus lowering the level of contamination during operations in ultra-high vacuum. In the development of bespoke positron image sensors, several features could be fine-tuned to further improve the performance in specific applications. Firstly, increasing the dimensions of the image sensor can ease its use as a device for the detection and optimization of positron beams. Secondly, high-precision temporal tagging can be achieved with an integrated front-end that transmits only the locations of individual positrons, rather than the entire image, following designs that are common for larger particle detectors~\cite{NIMFACT}. Thirdly, a decrease in pixel size would increase the precision of position measurements. The minimum size of pixels is constrained by the need of including support electronics and electrical connections in the die; this can be circumvented by employing pixels with a high aspect ratio when exceptional measurement precision is required in a specific direction at the cost of lowering the resolution in the orthogonal. Finally, bringing the active area closer to the surface, by reducing or omitting the passivation layer, would allow the detection of positrons at considerably lower energies and further compress the OTF of the sensor. This last improvement might prove challenging as handling a sensor with a shallow or no passivation layer necessitates particular care due to its sensitivity to atmospheric oxidants.

Positron detection via CMOS image sensors is a valuable technique that has the potential to impact a wide range of applications, from beam diagnostics, aiding the optimization of beam position, size and shape, to the enabling of fundamental physics experiments that are based upon the extremely accurate determination of positron positions. In particular, the precise readout of the positron impacting coordinates is crucial for high-precision diffractometry, which could allow the measurement of free-fall of antimatter~\cite{Mills_FreeFall}, or interferometric measurements, to test the quantum behaviour of positrons~\cite{Sala_Interferometry}. As a last example, this technology could enable the miniaturization and integration of transmission positron microscopes, as firstly demonstrated by Mills et al.~\cite{Mills_Transmission}, into experimental setups not specifically designed for this purpose. 

\section{Methods} \label{methods}

\subsection{Sample preparation} \label{MethodsSample}

To allow the Sony IMX219 image sensor to work with positrons, its color filter and microlens array need to be removed.
Prior to the removal procedure we prepared the sensors by sealing the bonding wires connecting the silicon die to the supporting PCB: we encased them in a protective layer of two component epoxy (UHU Plus Endfest 300) to shield them against accidental acid spills or mechanical impacts.
We then weakened the polymer layers by repeatedly dispensing small amounts of fuming nitric acid onto the sensor surface with a pipette and subsequently rinsing the device with deionized water and isopropyl alcohol. 
After repeatedly exposing the sensor to the acid, we mechanically removed the weakened layers by carefully scraping the surface using a custom-made tool, consisting of an aluminum scalpel with a flat tip, \SI{3}{\milli m} in width, sharpened at a $25^\circ$ angle.
After the scraping, we rinsed the sensor once again with isopropyl alcohol and tested it to verify that its imaging capabilities were not damaged during the procedure. 

At this point, the sensor can either be used as is -- exposing the bare silicon structure to the particle beam -- or be coated by a thin layer of high-efficiency phosphor, thus enabling the detection of very low energy positron beams.
We employed silver-doped zinc sulfide, denoted as ZnS:Ag (EJ-600) as our phosphor of choice. To obtain a uniform coating of the sensor, we suspended the powdered phosphor, \SI{8}{\micro\meter} in nominal particle size, in ethanol and then drop-wise deposited the suspension onto the sensor using a pipette. After the solvent has evaporated, the phosphor particles form an even coating of constant thickness on the sensor. We assessed the thickness of the coating by means of depth-of-focus measurements with an optical microscope. We then varied the amount and concentration of suspension used until we obtained a coating with a thickness of \SI{70}{\micro\meter}, chosen to mirror the thickness employed in literature for this type of phosphor~\cite{Stenson_Luminescence}.

\subsection{Readout electronics frontend}
\label{MethodsElectronics}

The sensor readout system consists primarily of a single-board computer (SBC, Raspberry 3B+) installed in the vacuum chamber. The SBC is powered via two HV feedthroughs which provide the acceleration bias voltage, which is connected to the SBC ground plane, and a high current \SI{5}{V} rail on top of the bias voltage which is used to power the SBC. The camera is installed onto a small daughter board and powered through the dedicated zero insertion force (ZIF) connector of the SBC. Data is transmitted to and from the SBC through an ad-hoc WiFi bridge. 

All of the electronics installed in vacuum were encased in a \SI{1.5}{\milli m} thick nonsealed copper shell which serves both as Faraday cage to capture potential arcs and as pathway to transfer heat to the passthroughs. We employed passthroughs with a copper core \SI{10}{\milli m} in diameter so that they could serve as the main heat transfer from the SBC to the environment. An additional copper brace was installed contacting the surface of the CPU with the copper casing to aid heat transfer from the system. Active air cooling had to be applied to the flange hosting the HV passthroughs and the \SI{5}{V} power supply to allow continuous operation of the board and sensor below $47^\circ$C.

The number of electronic components in vacuum was minimized and as many components as possible were removed from the SBC, including all of the connectors with the exception of the camera-dedicated ZIF. No electronic component containing liquids was installed in vacuum. Prior to the installation in vacuum all electronics were repeatedly washed with a cyclohexane-based cleaning agent (Electrolube Fluxclene FLU400DB) and rinsed with isopropyl alcohol before the evaporation of the cleaning agent to ensure the removal of any flux residuals.
The resulting assembly was able to be kept at a pressure of $2 \cdot 10^{-7}\SI{}{\milli bar}$ by the operation of a turbomolecular pump (Leybold TurboVac 300).

During the measurements, the sensitivity of the CMOS sensor was adjusted based on the intensity of the positron flux to minimize pile-ups, keep the signal within the dynamic range and maximize the S/N ratio.
We controlled the sensitivity of the CMOS by manipulating the sensor gain and exposure time which are determined by the values written in a variety of control registers.
When needed we have allowed the values of these registers to be set outside of the manufacturer's specifications~\cite{BergholdThesis}.

As the sensor sensitivity is proportional to the product of gain and exposure time, multiple configurations can result in the same sensitivity. We prioritized configurations that avoid using extreme gain or exposure values as we observed them to lead to increased fixed-pattern noise~\cite{BergholdThesis}; then determined the optimal configuration by dynamically adjusting the sensor parameters during the course of our measurements while monitoring the quality of the recorded images.
We provide a table in the supplementary materials, listing all of the register configurations used to acquire the data employed in this work.

\subsection{Beam fitting}
\label{MethodsFit}

Total and coincidence particle counts from the HPGe detectors allow us to measure the total positron flux impinging on the target. To compute from this the positron flux impinging on the detector we need to first determine the beam shape and position relative to the detector surface, this is due to the fact that the beam is typically not entirely contained inside the sensitive area of the sensor. To this end we model the beam profile with a two-dimensional Gaussian distribution which we write as
\begin{align}
    f_\xi(x,y) = A \cdot \exp\!\left[
            \frac{\vec{v}^{\,t} \cdot \Sigma \cdot \vec{v}}
                 {2\, (S_{xx}S_{yy} - S_{xy}^2)}%
        \right] \text{,}\label{BeamModel}
\end{align}
where
\begin{align*}
    \vec{v} &= \begin{bmatrix} x - C_x\\ y - C_y \end{bmatrix} \text{,}~~~~ 
    \Sigma   = \begin{bmatrix} S_{xx} & S_{xy} \\ S_{xy} & S_{yy} \end{bmatrix} \text{,}\\[2mm]
    \text{and} ~~~~ \xi     &= ( A,C_x,C_y,S_{xx},S_{xy},S_{yy} ) .
\end{align*}

We fit the beam by performing a least squares minimization combined with iterative reweighting (IRLS)~\cite{IRLS} and graduated optimization~\cite{GraduatedOptimization}. The use of IRLS reduces the likelihood of the minimization to be confounded by dead pixels or bright spots created by Compton processes, while graduated optimization prevents the fit from converging to a local minimum. We start the fit by subtracting the fixed-pattern noise, which is determined by acquiring a frame with the same setting used in the measurement but no positron beam.

The IRLS fit is performed via minimization for $k \in \{0,1,2\}$ of
\begin{equation}
    \chi^2(\xi_{k}) = \sum_{x,y} \frac{[f_{\xi_k}(x,y) - I(x,y)]^2}{w_k(x,y)} ~ \text{,}
    \label{ChiSquare}
\end{equation}
where $I(x,y)$ is the intensity of the measured image at the coordinates $(x,y)$ and $w_k(x,y)$ is recursively defined as
\begin{align}
    w_0(x,y)     &= 1 ~ \text{,} \nonumber     \\%
    w_{k+1}(x,y) &= \min(\delta, \lvert f_{\xi_k}(x,y) - I(x,y)\rvert) ~ \text{,}%
\end{align}
where $\delta$ is set to the value of an ADC unit. For each value of $k$ the minimization over the six free parameters $\xi$ is done by performing six iterations of the Newton method, with the gradient and Hessian matrix being computed analytically. The output of the IRLS fit is the value $\xi_2$.

Graduated optimization is used to fit each image, whereby Gaussian blurring is used to make the function to be optimized convex. The threefold IRLS minimization is repeated a total of $ \left \lfloor \log_4(W))\right \rfloor$ times, where $W$ is the width of the image in pixels. During the last iteration the original image is fitted; during the previous iterations a Gaussian blur with radius equal to $4^{i-1}$ pixels is applied to the image before the fit, with $i$ being the number of iterations left in the procedure. At each step the result of the previous fit is used to initialize the next one, which decreases the probability of the fit converging to a local minimum.
The procedure here described yields a function whose residuals, with reference to our experimental data, are observed to follow for the most part a normal distribution, with the dead pixels as outliers. The residual image has a uniform distribution both in the real and in the frequency domain and no distinguishable patterns, which brings confidence in the correctness of the fit model and technique.

\section*{Acknowledgments}
We would like to thank Dr.\ Davide Orsucci for insightful discussions and for reviewing the preliminary drafts of this paper.

\bibliography{bibliography}

\end{document}